# SOI RF SWITCH FOR WIRELESS SENSOR NETWORK

Wei Cai[1], Cheng Li[2] and ShiWei Luan[3]

[1]Department of Electrical Engineering and Computer Science, University of California, Irvine, USA
Caiw2@uci.edu

[2]Department of Computer Science, Rutgers University, New Brunswick, USA
chenglii@scarletmail.rutgers.edu

[3]Micro-Vu Corp, Santa Rosa, USA
shiwei.luan@gmail.com

*ABSTRACT*

*The objective of this research was to design a 0-5 GHz RF SOI switch, with 0.18um power Jazz SOI technology by using Cadence software, for health care applications. This paper introduces the design of a RF switch implemented in shunt-series topology. An insertion loss of 0.906 dB and an isolation of 30.95 dB were obtained at 5 GHz. The switch also achieved a third order distortion of 53.05 dBm and 1 dB compression point reached 50.06dBm. The RF switch performance meets the desired specification requirements.*

## 1. INTRODUCTION

Wireless medical sensor networks have offered significant improvements to all area, especially the healthcare industry in the 21st century [1][2]. Medical devices are attached on a patient's body, thus they can be directly monitor the physiological condition of patients. From the fundamental of the material to the highest architecture level, researches are working super hard, and the improvements can be seen here [3][4][5][6][7][8][9][10].

Wireless Sensor Networks (WSN) are extensively applied to solve a enormous array of problems at diverse conditions, such as to study nearly-extinguished animal's habits, forest fire detection [11][12][13][14][15][16]. When WSNs are widely spatially distributing to monitor patients, it can present nonstop, near-real time data over a huge residents [17][18][19][20]. WSNs can offer significant efficiencies compare to other solutions. For example, outpatient monitoring carries considerable cost, especially if applied to a large segment of the patient population. Possible cost-reduction solutions can be built by leveraging WSNs as well as cellular communication infrastructure. Industry, national research and academic are continue conducting frameworks [21][22][23][24][25][26][27][28].

Owing to existing technology limitations, WSNs of the health care are still in the early development stages [29][30][31]. Medial devices require Food and Drug

Administration (FDA) approval once it needs to successfully launching the market, this process would be time consuming and also challenge. So many research related to the material, device, components, architecture are developing quickly [31][32] [33][34]. Sensor nodes could be made up of WSNs, which contains a number of micro hardware components, such as MEMS sensors, batteries, IC chips, memory, antennas, etc [11][12][35][37][38].

## 2. BACKGROUND

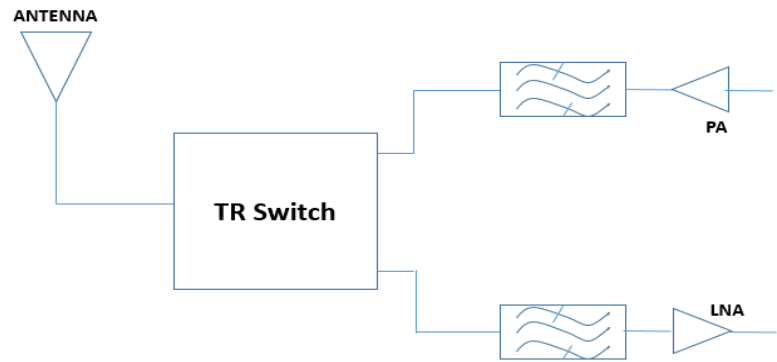

Figure 1. Block diagram of a multiband RF transceiver

The receiver will receive the signal and also has the digital signal processing to smooth out the data from the transmitter [37]. Figure 1 is the transceiver diagram. Due to the health care requiments, transmitter and receiver are both demanding low power. Director-conversion transmitter is very popular for such applications, because it offers good performance but low power as well as low cost.

This paper introduces the RF switch design, but the whole system, other blocks of the system such as PA, LNA, mixer are discussed in the paper [40][41][42]. In order to meet the standards, the RF switch is designed as shown in Table 1.

Table 1.  RF switch design requirement.

| Parameter | Size (Unit) |
|---|---|
| Frequency | 0-5G |
| Insolation loss (dB) | 0-1.2 (dB) |
| Isolation (dB) | >30 dB |
| IIP3 (dBm) | 50 dBm |
| P1dB (dBm) | 50 dBm |

FET switches usually have three different topologies, such as series, shunt and combinatorial topology. Due to modern, complicated requirements, users for in health care usually require the combinatorial topology to meet their stringent requirements, as seen in the Figure 3. When a control voltage is set high, the series FET would be on, which means a signal would pass to the following transistor, where a shunt FET would connect to the ground. When the control is set low, the series FET is off, so there will be no signal flow through the transistor, however the shunt FET will pass the signal.

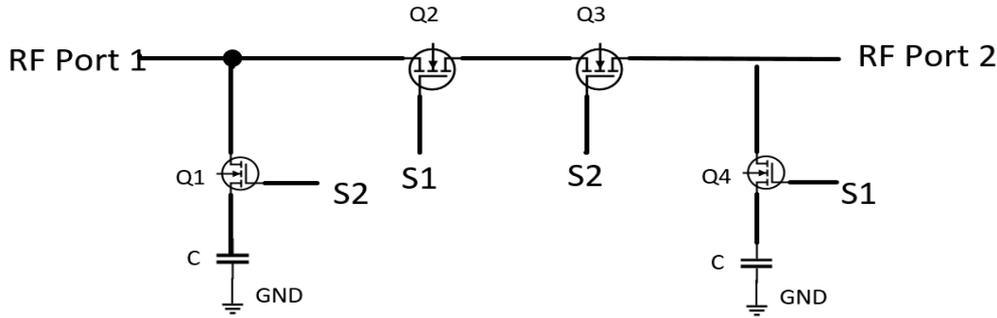

Figure 2. Combinational series-shunt FET switch

FETs are three terminal devices are usually fabricated as SOI or GaAs. The basic function is shown in Figure 2[32]. When the gate is more positively biased then the source, a channel would be formed between the source and drain side, so the resistance is lowered, and a current can be flowed between source and drain terminals. However, if the gate voltage is equal or smaller than the source terminal, no channel will form, resistance will be much higher, and no current flows through channel. The gates were controlled by the external digital signal, which is used "s" in the Figure 3. To turn-on the switch, setting "s" to 2.4 V, then the switch is at on stage. When S is set to 0V, the switch is at off stage.

In terms of RF switch performance, there are several key parameters, such as reflection coefficient S11, insertion loss S21 and isolation S31[32].

S11 is the input reflection coefficient, which is voltage ratio of the reflected wave on the input port to the original wave. This parameter represents the power loss from impedance mismatches, also known as voltage standing wave ratio (VSWR).

S21 represents the forward voltage gain. A low insertion loss between source and active switch is critical to increase the efficiency.

S31 is also a very important switch parameter. When there is 3 ports, two ports are on, and another port is off, and this parameter is a measure of the transmission coefficient from the source to the off arm. This parameter represents how much power was leaked into the off arm.

Besides the S11, S21, S31, a switch design's value must also consider the intercept point (IP3). This parameter is a measure of the linearity of a device, which also known as intermodulation distortion. The third order intercept point is where the intercept of the fundamental frequency and the third order of the fundamental frequency.

## 3. DESIGN

The SPST switch was used for the healthcare application which can worked for the 0-5 GHz, so it required designed high isolation and moderate insertion loss. A proposed topology is shown in Figure 3, and which is based on shunt-series configuration [40]. Such type of topology slightly different from Figure 2 when only using a single series-shunt topology [40]. In Figure 2, only four FETs are shown. For many commercial switches, however, each of these FETs would represent a stack of at least twelve devices in series. Stacking the devices in this manner is often necessary since each individual FET has a relatively low breakdown voltage, as previously discussed. By stacking the FETs, the relatively high voltage typically used in most RF frontends is dispersed over several devices, so that the voltage across any one device is relatively small and unlikely to result in breakdown. Ideally, all the FETs in the stack should have identical drain-to-source voltages.

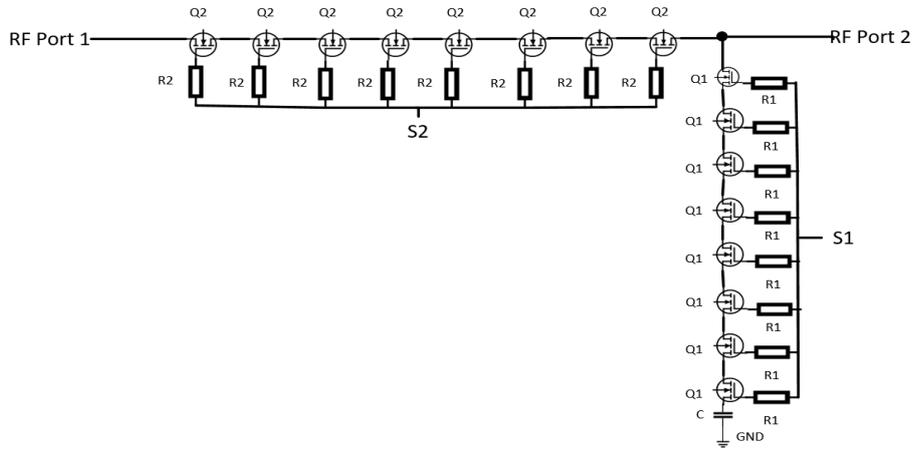

Figure 3. Modified series-shunt FET switch

Table 2. RF switch component.

| Parameter | Size (Unit) |
|---|---|
| Q1 | W/L=10um/0.48um (f=60,m=40) |
| Q2 | W/L=10um/0.58um (f=20,m=40) |

## 4. RESULTS

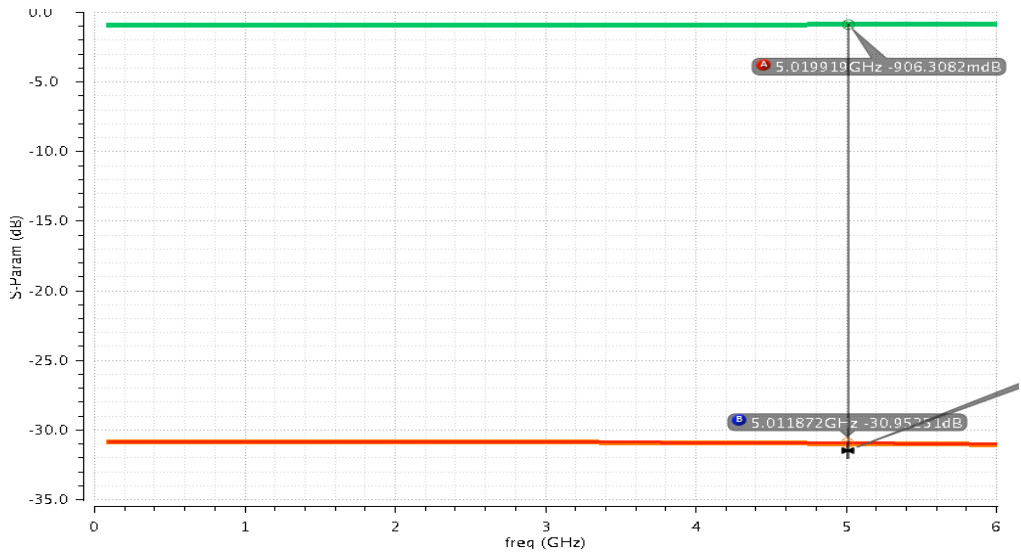

Figure 4. Insertion (green) and isolation (red)

As seen in Figure 4, an insertion loss of 0.906 dB and an isolation of 30.95 dB were obtained at 5 GHz.

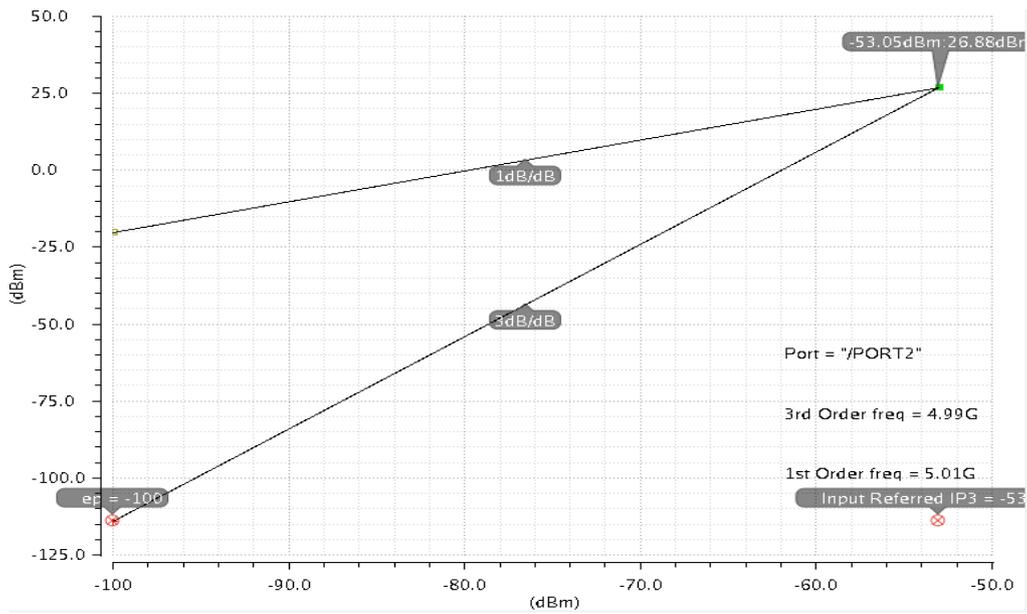

Figure 5. IIP3

As seen in Figure 5, input referred IP3 reached 53.05 dBm.

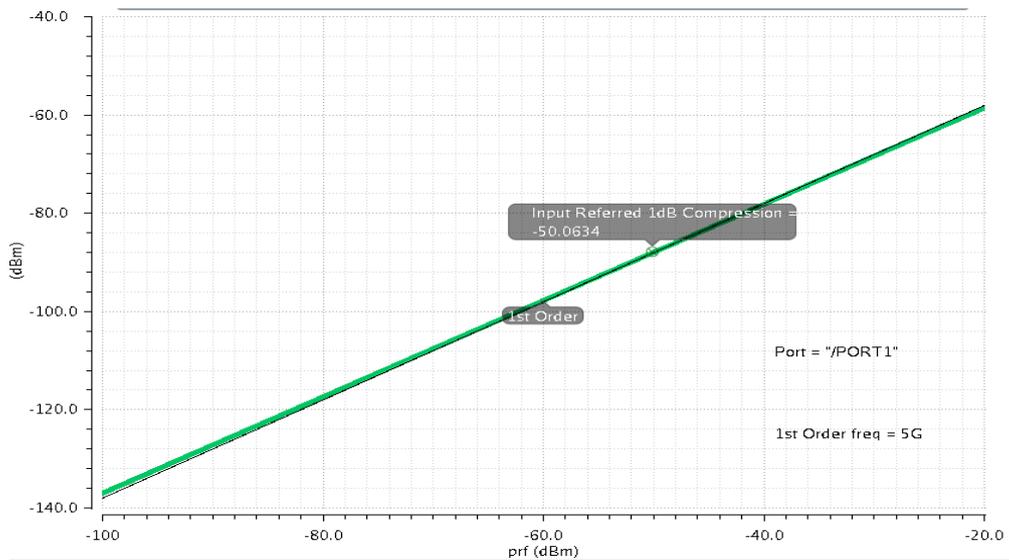

Figure 6. P1dB

As seen in Figure 6, 1 dB compression point reached the switch achieved a third order distortion of 50.06 dBm.

As seen in Figure 7, by adding the impedance matching circuit, S11 could get below 10 dB.

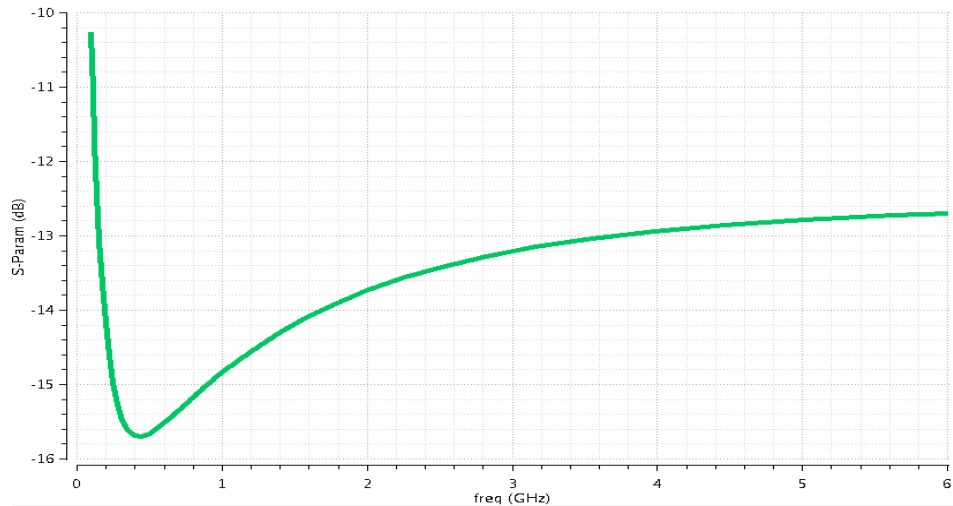

Figure 7. S11

## 5. CONCLUSION

Simulating RF switch using cadence software based on Power Jazz SOI process 180nm technology is presented by this paper. This RF switch of the health care application is positively to be realized, but it still has extra room to improve if more requirements added. Communication could be by the medical sensors; such technology could help to link between the patients and doctors. Hospitalization resources could be improved by implementing low cost and low power medical devices. To realize this, future improvement is needed.

**Authors**

Wei Cai is a graduate student at the University of California, Irvine, CA. She received her Masters degree from Dept. of Electrical Engineering, University of Hawaii at Manoa and Bachelor degree from Zhejiang University, China. Her research interests include device physics simulation, analog/ RF circuit design.

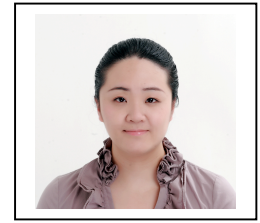

Cheng Li received the PhD degree in Computer Science from Rutgers University. His research interests include big data, performance management, high performance computing and Internet services.

Shiwei Luan received his Master degree in Electrical and Computer Engineering from Kansas State University. He is working as an Electrical Engineer specialized on schematic capture and PCB layout design of high speed digital system with CPU, memory, USB, PCIe, and Ethernet